## Chapter 6

## Cosmic polarization rotation: An astrophysical test of fundamental physics[*]

Sperello di Serego Alighieri

*INAF - Osservatorio Astrofisico di Arcetri,*
*Largo E. Fermi 5, 50125 Firenze, Italy*
*sperello@arcetri.astro.it*

Possible violations of fundamental physical principles, e.g. the Einstein equivalence principle on which all metric theories of gravity are based, including general relativity (GR), would lead to a rotation of the plane of polarization for linearly polarized radiation traveling over cosmological distances, the so-called cosmic polarization rotation (CPR). We review here the astrophysical tests which have been carried out so far to check if CPR exists. These are using the radio and ultraviolet polarization of radio galaxies and the polarization of the cosmic microwave background (both E-mode and B-mode). These tests so far have been negative, leading to upper limits of the order of one degree on any CPR angle, thereby increasing our confidence in those physical principles, including GR. We also discuss future prospects in detecting CPR or improving the constraints on it.

### 1. Introduction

Linear polarization is a simple phenomenon by which a single photon is able to transmit across the universe the information about the orientation of a plane. The question which we discuss in this paper is whether the orientation of the plane of linear polarization, the so-called position angle (PA[a]), is conserved for electromagnetic radiation traveling long distances, i.e. if there is any cosmic polarization rotation (CPR). Clearly, if the CPR angle $\alpha$ is not zero, symmetry must be broken at some level, since $\alpha$ must be either positive or negative, for a counterclockwise or clockwise rotation. This immediately suggests that CPR should be connected with the violation of fundamental physical principles. Indeed, it is linked also to a possible violation of the Einstein equivalence principle (EEP), which is the foundation of any metric theory of gravity, including general relativity (GR). Therefore it deserves a chapter in this volume.

---

[*]This article was also published in *Int. J. Mod. Phys. D* **24**, 1530016 (2015). This version has been updated to include the results of Planck and POLARBEAR.
[a]We adopt the International Astronomical Union (IAU) convention for PA: it increases counterclockwise facing the source, from North through East.[36]

The fundamental principles whose violation would imply CPR are briefly discussed in Sec. 2 (please refer also to other chapters in this volume). For most of them, the CPR angle would be independent of wavelength. However the violation of some principles would imply a wavelength-dependent CPR, not to be confused with the Faraday rotation, which is a well-known effect for radiation passing through a plasma with a magnetic field. CPR, if it exists, would occur in vacuum. CPR has sometimes been inappropriately called "cosmological birefringence." However we follow here the advice of Ni,[60] since birefringence is only appropriate for a medium whose index of refraction depends on the direction of polarization of the incident light beam, which is then split in two. The phenomenon we are considering here is pure rotation of the polarization, without any splitting.

Testing for CPR is simple in principle: it requires a distant source of linearly polarized radiation, for which the orientation $PA_{em}$ of the polarization at the emission can be established. Then CPR is tested by comparing the observed orientation $PA_{obs}$ with $PA_{em}$:

$$\alpha = PA_{obs} - PA_{em}.$$

In practice, it is not easy to know *a priori* the orientation of the polarization for a distant source: in this respect the fact that scattered radiation is polarized perpendicularly to the plane containing the incident and scattered rays has been of great help, applied both to radio galaxies (RGs) (see Sec. 4) and to the cosmic background (CMB) radiation (see Sec. 5). For those cases in which CPR depends on wavelength, one can also test CPR by simply searching for variation of PA with the wavelength of the radiation, even without knowing $PA_{em}$. In this paper, we will review the astrophysical methods which have been used to test CPR, we list the results of these test, discuss the advantages and disadvantages of the various methods and suggest future prospects for these tests.

## 2. Impact of CPR on Fundamental Physics

This possibility of CPR arises in a variety of important contexts, like the presence of a cosmological pseudoscalar condensate, Lorentz invariance violation and charge, parity and time reversal (CPT) violation, neutrino number asymmetry, the EEP violation. In particular, the connection of the latter with CPR is relevant for this GR Centennial year, since all metric theories of gravity, including GR are based on the EEP. Since the weak equivalence principle (WEP) is tested to a much higher accuracy than the EEP, Schiff[68] conjectured that any consistent Lorentz-invariant theory of gravity which obeys the WEP would necessarily also obey the EEP. If these were true, the EEP would tested to the same accuracy as the WEP, increasing our experimental confidence in GR. However, Ni[57,58] found a unique counter example to Schiff's conjecture: a pseudoscalar field which would lead to a violation of the EEP, while obeying the WEP. Such field would produce a CPR. Therefore, testing for the CPR is important for our confidence in GR. For the other theoretical impacts of CPR we refer the reader to Refs. 59, 60 and 62.

### 3. Constraints from the Radio Polarization of RGs

Already in his seminal paper about the unique counter-example to Schiff's conjecture giving rise to CPR, Ni[57] suggested that observations of polarized astrophysical sources could give constraints on the CPR. However, only in 1990, the polarization at radio wavelengths of RGs and quasars was used for the first astrophysical test of CPR.[12,b] Ref. 12 has used the fact that extended radio sources, in particular, the more strongly polarized ones, tend to have their plane of integrated radio polarization, corrected for Faraday rotation, usually perpendicular and occasionally parallel to the radio source axis,[18] to put a limit of 6° at the 95% confidence level (CL) to any rotation of the plane of polarization for the radiation coming from these sources in the redshift interval $0.4 < z < 1.5$.

Reanalyzing the same data, Nodland and Ralston[63] claimed to have found a rotation of the plane of polarization, independent of the Faraday one, and correlated with the angular positions and distances to the sources. Such rotation would be as much as 3 rad for the most distant sources. However, several authors have independently and convincingly rejected this claim, both for problems with the statistical methods,[13,27,51] and by showing that the claimed rotation is not observed for the optical/ultraviolet (UV) polarization of two RGs (see below) and for the radio polarization of several newly observed RGs and quasars.[72]

In fact, the analysis of Leahy[48] is important also because it introduces a significant improvement to the radio polarization method for the CPR test. The problem with this method is the difficulty in estimating the direction of the polarization at the emission. Since the radio emission in RGs and quasars is due to synchrotron radiation, the alignment of its polarization with the radio axis implies an alignment of the magnetic field, which is not obvious *per se*. In fact, theory and magnetohydrodynamics simulations foresee that the projected magnetic field should be perpendicular to strong gradients in the total radio intensity.[7,66] For example, for a jet of relativistic electrons the magnetic field should be perpendicular to the local jet direction at the edges of the jet and parallel to it where the intensity changes along the jet axis.[10] On the other hand, such alignments are much less clear for the *integrated* polarization, because of bends in the jets and because intensity gradients can have any direction in the radio lobes, which emit a large fraction of the polarized radiation in many sources. In fact, it is well-known that the peaks at 90° and 0° in the distribution of the angle between the direction of the radio polarization and that of the radio axis are very broad and the alignments hold only statistically, but not necessarily for individual sources (see e.g. Fig. 1 of Ref. 12). More stringent tests can be carried out using high angular resolution data on radio polarization and the local magnetic field's alignment for individual sources,[72] although to our knowledge, only once[48] this method has been used to put quantitative limits on the polarization rotation. For example, Carroll,[14] using the data on the ten RGs of Leahy,[48] obtains an average constraint on any CPR angle of $\alpha = -0.6° \pm 1.5°$ at the mean redshift $\langle z \rangle = 0.78$. However, the preprint by Leahy[48] remained unpublished and does not explain convincingly how the angle between the direction of the local intensity gradient and that of the polarization is derived. For example, for 3C9, the source with the best accuracy, Leahy[48] refers to Ref. 47, who however, do not give any measurements of local gradients.

---

[b]Ref. 9 had earlier claimed a substantial anisotropy in the angle between the direction of the radio axis and the direction of linear radio polarization in a sample of high-luminosity classical double radio sources, but used it to infer rotation of the universe, not to test for CPR.

### 4. Constraints from the UV Polarization of RGs

Another method to test for CPR has used the perpendicularity between the direction of the elongated structure in the UV[c] and the direction of linear UV polarization in distant powerful RGs. The test was first performed by Refs. 16 and 24, who obtained that any rotation of the plane of linear polarization for a dozen RGs at $0.5 \leq z \leq 2.63$ is smaller than $10°$.

Although this UV test has sometimes been confused with the one at radio wavelengths, probably because they both use RGs polarization, it is a completely different and independent test, which hinges on the well-established unification scheme for powerful radio-loud Active Galatic Nuclei (AGN).[5] This scheme foresees that powerful radio sources do not emit isotropically, but their strong UV radiation is emitted in two opposite cones, because the bright nucleus is surrounded by an obscuring torus: if our line of sight is within the cones, we see a quasar, otherwise we see a RG. Therefore, powerful RGs have a quasar in their nuclei, which can only be seen as light scattered by the interstellar medium of the galaxy. Often, particularly in the UV, this scattered light dominates the extended radiation from RGs, which then appear elongated in the direction of the cones and strongly polarized in the perpendicular direction.[23] The axis of the UV elongation must be perpendicular to the direction of linear polarization, because of the scattering mechanism which produces the polarization. Therefore, in this case it is possible to accurately predict the direction of polarization at the emission and compare it with the observed one. This method of measuring the polarization rotation can be applied to any single case of distant RG, which is strongly polarized in the UV, allowing independent CPR tests in many different directions. Another advantage of this method is that it does not require any correction for Faraday rotation, which is large at radio wavelengths, but negligible in the UV.

In the case of well resolved sources, the method can be applied also to the polarization which is measured locally at any position in the elongated structures around RGs, and which has to be perpendicular to the vector joining the observed position with the nucleus. From the polarization map in the V-band ($\sim 3000\,\text{Å}$ rest-frame) of 3C 265, a RG at $z = 0.811$,[70] the mean deviation of the 53 polarization vectors plotted in the map from the perpendicular to a line joining each to the nucleus is $-1.4° \pm 1.1°$.[72] However, more distant RGs are so faint that only the integrated polarization can be measured, even with the largest current telescopes: strict perpendicularity is expected also in this case, if the extended emission is dominated by the scattered radiation, as is the case in the UV for the strongly polarized RGs.[71]

---

[c]When a distant RG ($z > 0.7$) is observed at optical wavelengths ($\lambda_{\text{obs.}} \sim 5000\,\text{Å}$), these correspond to the UV in the rest frame ($\lambda_{\text{em.}} \leq 3000\,\text{Å}$).

Recently, the available data on all RGs with redshift larger than two and with the measured degree of linear polarization larger than 5% in the UV (at $\sim 1300$ Å) have been reexamined, and no rotation within a few degrees in the polarization for any of these eight RGs has been found.[25] In addition, assuming that the CPR angle should be the same in every direction, an average constraint on this rotation $\langle \alpha \rangle = -0.8° \pm 2.2°$ $(1\sigma)$ at the mean redshift $\langle z \rangle = 2.80$ has been obtained.[25] The same data have been used by Ref. 39 to set a CPR constraint in case of a nonuniform polarization rotation, i.e. a rotation which is not the same in every direction: in this case the variance of any rotation must be $\langle \alpha^2 \rangle \leq (3.7°)^2$. The CPR test using the UV polarization has advantages over the other tests at radio or CMB wavelengths, if CPR effects grow with photon energy (the contrary of Faraday rotation), as in a formalism where Lorentz invariance is violated but CPT is conserved.[43,44]

## 5. Constraints from the Polarization of the CMB Radiation

A more recent method to test for the existence of CPR is the one that uses the CMB polarization, which is induced by the last Thomson scattering of decoupling photons at $z \sim 1100$, resulting in a correlation between temperature gradients and polarization.[49] CMB photons are strongly linearly polarized, since they result from scattering. However the high uniformity of CMB produces a very effective averaging of the polarization in any direction. It is only at the CMB temperature disuniformities that the polarization does not average out completely and residual polarization perpendicular to the temperature gradients is expected. Therefore, also for the CMB polarization it is possible to precisely predict the polarization direction at the emission and to test for any CPR. After the first detection of CMB polarization anisotropies by Degree Angular Scale Interferometer (DASI),[46] there have been several CPR tests using the CMB E-mode polarization pattern.

Unfortunately, the scientists working on the CMB polarization have adopted for the polarization angle a convention which is opposite to the IAU one, used for decades by all other astrophysicists and enforced by the IAU[36]: for the CMB polarimetrists, following a software for the data pixelization on a sphere,[30] the polarization angle increases clockwise, instead of counterclockwise, facing the source. This produces an inversion of the U Stokes parameter, corresponding to a change of PA sign. Obviously, these different conventions have to be taken into account, when comparing data obtained with the different methods used for CPR searches. As mentioned in the introduction, all PA in this paper are given in the IAU convention. Independently of the adopted convention, a problem of CMB polarimetry is the calibration of the PA, which is not easy at CMB frequencies. Although different methods are used, like the *a priori* knowledge of the detector's orientation, the use of calibration sources both near the experiment on the ground and on the sky, the current calibration accuracy is of the order of one degree, producing a nonnegligible systematic error $\beta$ on the measured PA. In order to alleviate the PA calibration problem, Ref. 42 have suggested a self-calibration technique consisting in minimizing EB and TB power spectra with respect to PA offset. Unfortunately, such a calibration technique would eliminate not just the PA calibration offset $\beta$, but also $\alpha - \beta$, where $\alpha$ is the uniform CPR angle, if it exists. Therefore, no independent information on the uniform CPR angle can be obtained, if this calibration technique is adopted, like with the Background Imaging of Cosmic Extragalactic Polarization 2 (BICEP2)[2] experiment.

In the following, we summarize the most recent and accurate CPR measurements obtained using the CMB polarization (see Table 1 and Fig. 1). The BOOMERanG collaboration, revisiting the limit set from their 2003 flight, found a CPR angle $\alpha = 4.3° \pm 4.1°$ (68% CL), assuming uniformity over the whole sky.[64] The QUEST at DASI (QUaD) collaboration found $\alpha = -0.64° \pm 0.50°$ (stat.) $\pm 0.50°$ (syst.) (68% CL),[11] while using three years of BICEP1 data one gets $\alpha = 2.77° \pm 0.86°$ (stat.) $\pm 1.3°$ (syst.) (68% CL).[40] Combining nine years of Wilkinson microwave anisotropy probe (WMAP) data and assuming uniformity, a limit to CPR angle $\alpha = 0.36° \pm 1.24°$ (stat.) $\pm 1.5°$ (syst.) (68% CL) has been set, or $-3.53° < \alpha < 4.25°$ (95% CL), adding in quadrature statistical and systematic errors.[33] The POLARBEAR collaboration[3] reports about a difference of 1.08° in the instrument polarization angle obtained at 148 GHz minimizing the EB spectrum and that obtained from their data on the Crab Nebula using the PA measurement at 90 GHz of Ref. 6. This corresponds to a measurement of CPR, performed with the effect of a rotation on the EB spectrum and using the Crab Nebula for the PA calibration, and giving a CPR angle $\alpha = 1.08° \pm 0.2°$(stat.) $\pm 0.5°$(syst.), assuming that the Crab Nebula polarization angle does not change between 90 and 148 GHz. A consistency check with the value of the Cen A polarization angle measured by POLARBEAR confirms this result. Recently the ACTPol (Atacama Cosmology Telescope Polarimeter) team[56] have used their first three months of observations to measure the CMB polarization over four sky regions near the celestial equator. They do not give an explicit value for the CPR, also because they have used the EB and TB[d] power minimization technique of Ref. 42. However it is possible to derive a value of the CPR from their data, since they have measured a PA of $150.9 \pm 0.6°$ for Crab Nebula (Tau A, a polarization standard source), using the EB and TB nulling procedure (Hasselfield, private communication). The most precise fiducial measurement at CMB frequencies of the Crab Nebula polarization angle is a PA of $149.9 \pm 0.2°$ at 90 GHz.[6] If we assume that the Crab Nebula polarization PA would not change

---

[d]EB and TB are the cross-correlation power spectra between E- and B-modes and between temperature and B-mode.

Table 1. Measurements of CPR with different methods (in chronological order).

| Method | CPR angle ± stat. (± syst.) | Frequency or $\lambda$ | Distance | Direction | Reference |
|---|---|---|---|---|---|
| RG radio pol. | $|\alpha| < 6°$ | 5 GHz | $0.4 < z < 1.5$ | All-sky (uniformity ass.) | 12 |
| RG UV pol. | $|\alpha| < 10°$ | $\sim 3000$ Å rest-frame | $0.5 < z < 2.63$ | All-sky (uniformity ass.) | 16,17 |
| RG UV pol. | $\alpha = -1.4° \pm 1.1°$ | $\sim 3000$ Å rest-frame | $z = 0.811$ | $RA : 176.4°, Dec : 31.6°$ | 72 |
| RG radio pol. | $\alpha = -0.6° \pm 1.5°$ | 3.6 cm | $\langle z \rangle = 0.78$ | All-sky (uniformity ass.) | 14,48 |
| CMB pol. BOOMERanG | $\alpha = 4.3° \pm 4.1°$ | 145 GHz | $z \sim 1100$ | $RA \sim 82°, Dec \sim 45°$ | 64 |
| CMB pol. QUAD | $\alpha = -0.64° \pm 0.50° \pm 0.50°$ | 100–150 GHz | $z \sim 1100$ | $RA \sim 82°, Dec \sim 50°$ | 11 |
| RG UV pol. | $\alpha = -0.8° \pm 2.2°$ | $\sim 1300$ Å rest-frame | $\langle z \rangle = 2.80$ | All-sky (uniformity ass.) | 25 |
| RG UV pol. | $\langle \delta\alpha^2 \rangle \leq (3.7°)^2$ | $\sim 1300$ Å rest-frame | $\langle z \rangle = 2.80$ | All-sky (stoch. var.) | 25,39 |
| CMB pol. WMAP9 | $\alpha = 0.36° \pm 1.24° \pm 1.5°$ | 23–94 GHz | $z \sim 1100$ | All-sky (uniformity ass.) | 33 |
| CMB pol. BICEP1 | $\alpha = 2.77° \pm 0.86° \pm 1.3°$ | 100–150 GHz | $z \sim 1100$ | $-50° < RA < 50°, -70° < Dec < -45°$ | 40 |
| CMB pol. POLARBEAR | $\alpha = 1.08° \pm 0.2° \pm 0.5°$ | 148 GHz | $z \sim 1100$ | $RA \sim 70°, 178°, 345°; Dec \sim -45°, 0°, -33°$ | 3 |
| CMB pol. ACTPol | $\alpha = 1.0° \pm 0.63°\,**$ | 146 GHz | $z \sim 1100$ | $RA \sim 35°, 150°, 175°, 355°, Dec \sim 50°$ | 54,56 |
| CMB pol. B-mode | $\langle \delta\alpha^2 \rangle \leq (1.36°)^2$ | 95–150 GHz | $z \sim 1100$ | Various sky regions | 54 |
| CMB pol. Planck | $\alpha = 0.35° \pm 0.05° \pm 0.28°$ | 30–353 GHz | $z \sim 1100$ | All-sky (uniformity ass.) | 79 |

*Note*: **A systematic error should be added, equal to the unknown difference of the Crab Nebula polarization PA between 146 GHz and 90 GHz.

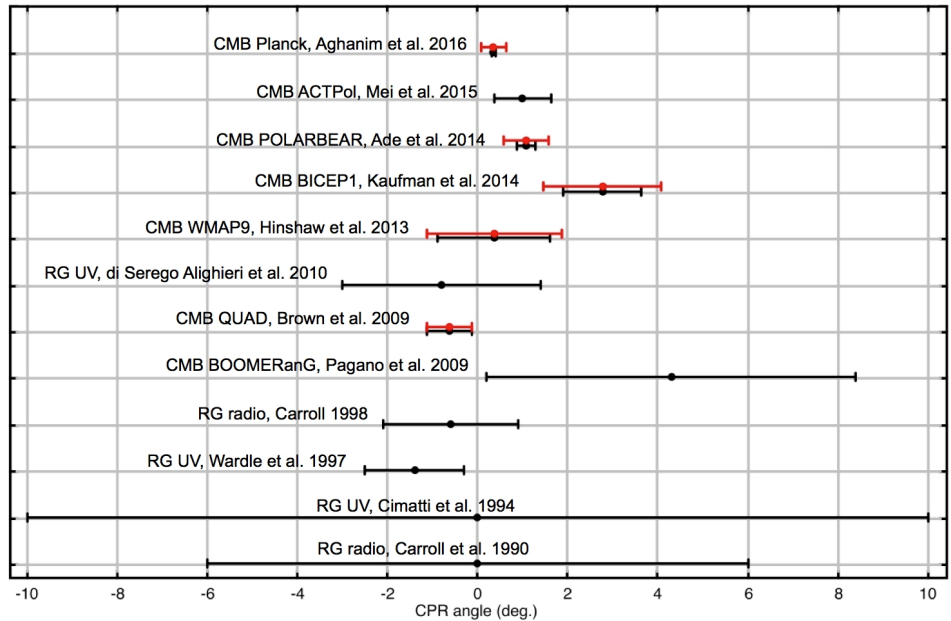

Fig. 1. CPR angle measurements by the various experiments, displayed in chronological order. Black error bars are for the statistical error, while red ones are for the systematic one, if present. A systematic error should be added to the ATCPol measurement, equal to the unknown difference of the Crab Nebula PA between 146 GHz and 90 GHz.[54]

between 90 GHz and 146 GHz (see e.g. the discussion in Sec. 6 of Ref. 6), then the average CPR angle over the ACTPol equatorial regions would be the difference between the above values $\alpha = 1.0° \pm 0.63°$ (see Ref. 54); however the above assumption leaves room for some systematic error. We could instead use the data of Ref. 56 in a different way: since the PA offset angle which they obtain from the EB minimization technique is $-0.2° \pm 0.5°$, i.e. consistent with zero, Ref. 56 suggests that their optical modeling procedure should be free of systematic errors at the 0.5° level or better. If these were true, then $\alpha = 0.22° \pm 0.32° \pm 0.5°$.[54] In summary, for the ACTPol result we prefer the assumption on the constancy of the Crab Nebula polarization angle between 90 GHz and 146 GHz, also because this can be tested *a posteriori* and an eventual correction applied. Recently the results on CPR from the Planck satellite have finally been published giving $\alpha = 0.35° \pm 0.05° \pm 0.28°$ with the stacking analysis.[78] Thanks to the very good quality of the Planck data, they achieve, as expected, a very small statistical uncertainty, considerably lower than any previous measurement. However their accuracy is limited by the uncertainty in the calibration of the position angle: even using the best calibrators, their systematic uncertainty is more than 5 times larger than the statistical one. In fact, most likely their measurement of the CPR angle (see Table 1) is actually a measurement of the Planck polarization angle offset.

In summary, although some have claimed to have detected a rotation,[40,75] the CMB polarization data appear well consistent with a null CPR. In principle the CMB polarization pattern can be used to test CPR in specific directions. However, because of the extremely small anisotropies in the CMB temperature and polarization, these tests have so far used averages over relatively large regions of sky, assuming uniformity.

Recently, Ref. 26 has suggested the possibility of setting constraints on the CPR also using measurements of the B-mode polarization of the CMB, because of the coupling from E-mode to B-mode polarization that any such rotation would produce. This possibility is presently limited by the relatively large systematic errors on the polarization angle still affecting current data. The result is that from the SPTpol (South Pole Telescope polarimeter), POLARBEAR and BICEP2 B-mode polarization data it is only possible to set constraints on the fluctuations $\langle \delta\alpha^2 \rangle \leq (1.56°)^2$ of the CPR, not on its mean value. Ref. 54 have similarly obtained an upper limit on the CPR fluctuations $\langle \delta\alpha^2 \rangle \leq (1.68°)^2$ from the ACTPol B-mode data of Ref. 56. By considering also SPTPol B-mode polarization data, Ref. 79 have recently improved this upper limit to $\langle \delta\alpha^2 \rangle \leq (0.97°)^2$. The one-but-last row of Table 1 reports the combined constraint on the CPR fluctuations obtained from all the B-mode data mentioned above.

## 6. Other Constraints

Observations of nearby polarized galactic objects could contribute to the CPR test, in particular, for those cases where polarization measurements can be made with high accuracy and at very high frequencies (useful if CPR effects grow with photon energy). Pulsars and supernova remnants emit polarized radiation in a very broad frequency range. For example, hard X-ray polarization observations of the Crab Nebula[22] have been used to set a limit to CPR angle $\alpha = -1° \pm 11°$.[52] However this limit is not particularly stringent, both because of the low accuracy of the X-ray polarization measurement and because of the limited distance to the source. In future, more precise X-ray polarization experiments such as POLARIX,[19] could much improve the situation.

Gamma-ray bursts (GRB) are very distant sources which emit polarized radiation both in the optical afterglow[20,73] and in the prompt gamma-ray emission.[31,38] Nevertheless, they cannot be used for CPR searches, since the orientation of the polarization at the emission is unknown. However, they can be used to test for birefringence effects, i.e. an energy-dependent rotation of the polarization angle, such as those produced by Lorentz invariance violation,[50] since the detection of linear polarization in a gamma-ray band excludes a significant rotation of the polarization within that energy band. In this way Ref. 31 was able to put an upper limit to the dimensionless parameter[e] of this birefringence effect of $\xi < 1 \times 10^{-16}$ from the gamma-ray polarization of a GRB at $z = 2.74$. Using the same data for testing the Lorentz symmetry and the equivalence principle, Refs. 61 and 62 provide a birefringence constraint of about $10^{-38}$.

---

[e]$\xi \equiv (n_0)^3$, where $n_0$ is the time component of the Myers–Pospelov four-vector $n_\alpha$, in a reference frame where $n_\alpha = (n_0, 0, 0, 0)$.[32,55]

For an issue related to CPR, Ref. 34 provides evidence that the directions of linear polarization at optical wavelengths for a sample of 355 quasars ($0 \leq z \leq 2.4$) are nonuniformly distributed, being systematically different near the North and South Galactic Poles, particularly in some redshift ranges. Such behavior could not be caused by uniform CPR, since a rotation of randomly distributed directions of polarization could produce the observed alignments only with a very contrived distribution of rotations as a function of distance and position in the sky. Moreover, the claim by Ref. 34 has not been confirmed by the radio polarization directions on a much larger sample of 4290 flat-spectrum radio sources,[37] and Ref. 35 recently suggested that the alignments could be due to an alignment of quasar's spin axis to the structures to which they belong. The possibility that the quasar's polarization alignments could be due to the mixing of photons with axion-like particles is excluded by the absence of circular polarization.[65]

The rotation of the plane of linear polarization can be seen as different propagation speeds for right and left circularly polarized photons ($\Delta c/c$). The sharpness of the pulses of pulsars in all Stokes parameters can be used to set limits corresponding to $\Delta c/c \leq 10^{-17}$. Similarly the very short duration of GRB gives limits of the order of $\Delta c/c \leq 10^{-21}$. However the lack of linear polarization rotation discussed in the previous sections can be used to set much tighter limits ($\Delta c/c \leq 10^{-32}$).[29]

In a complementary way to the astrophysical tests described in the previous sections, also laboratory experiments can be used to search for CPR. These are outside the scope of this paper and have not obtained significant constraints. For example, the PVLAS (Polarizzazione del Vuoto con LASer) collaboration has found a polarization rotation in the presence of a transverse magnetic field,[76] but later refuted this claim, attributing the rotation to an instrumental artifact.[77] The null result is consistent with the measurement of Ref. 15.

## 7. Discussion

Table 1 and Fig. 1 summarize the most important limits set on the CPR angle with the various methods examined in the previous sections. Only the best and most recent results obtained with each method are listed. For uniformity, all the results for the CPR angle are listed at the 68% CL ($1\sigma$), except for the first one, which is at the 95% CL, as given in the original Ref. 12. In general, all the results are consistent with each other and with a null CPR. Even the CMB measurement by BICEP1, which apparently shows a nonzero rotation at the $2\sigma$ level, cannot be taken as a firm CPR detection, since it has not been confirmed by other more accurate measurements.

In practice, all CPR test methods have reached so far an accuracy of the order of 1° and $3\sigma$ upper limits to any rotation of a few degrees. It has been however useful to use different methods since they are complementary in many ways. They cover different wavelength ranges and, although most CPR effects are wavelength independent, the methods at shorter wavelength have an advantage, if CPR effects grow with photon energy. They also reach different distances, and the CMB method is unbeatable in this respect. However the relative difference in light travel time between $z = 3$ and $z = 1100$ is only 16%. The radio polarization method, when it uses the integrated polarization, has the disadvantage of not relying on a firm prediction of the polarization orientation at the source, which the other methods have. In addition, the radio method requires correction for Faraday rotation. All methods can potentially test for a rotation which is not uniform in all directions, although this possibility has not yet been exploited by the CMB method, which also cannot see how an eventual rotation would depend on the distance. Reference 28 have recently examined the dependence of CPR on the wavelength and on the distance of the source, and found none, which is not surprising for a null (so far) CPR: in practice, they cannot improve the limit already set on the birefringence parameter $\xi$ in Ref. 31 (see Sec. 6).

## 8. Outlook

In the future, improvements can be expected for all methods, e.g. by better targeted high resolution radio polarization measurements of RGs and quasars, by more accurate UV polarization measurements of RGs with the coming generation of giant optical telescopes,[8,21,67] and by future CMB polarimeters such as BICEP3[4] and COrE+.[80] In any case, since at the moment the limiting factor for improving the constraints on the CPR angle with the CMB are the systematic uncertainty on the calibration of the polarization angle, it will be necessary to reduce these, which at the moment is at best $0.3°$ for CMB polarization experiments. The best prospects to achieve this improvement are likely to be more precise measurements of the polarization angle of celestial sources at CMB frequencies, e.g. with the Australia Telescope Compact Array[53] and with Atacama Large Millimeter/Submillimeter Array (ALMA),[69] and a calibration source on a satellite.[41]

## Acknowledgments

We would like to thank Matthew Hasselfied, Matteo Galaverni and Wei-Tou Ni for useful discussions.

## References

1. Planck Collab. (P. A. R. Ade *et al.*), *Astron. Astrophys.* **571** (2014) A16.
2. BICEP2 Collab. (P. A. R. Ade *et al.*), *Phys. Rev. Lett.* **112** (2014) 241101.
3. POLARBEAR Collab. (P. A. R. Ade *et al.*), *Astrophys. J.* **794** (2014) 171.
4. Z. Ahmed *et al.*, *Proc. SPIE-Int Soc. Opt. Eng.* **9153** (2014) 1.
5. R. Antonucci, *Annu. Rev. Astron. Astrophys.* **31** (1993) 473.
6. J. Aumont *et al.*, *Astron. Astrophys.* **514** (2010) A70.

7. M. C. Begelman, R. D. Blandford and M. J. Rees, *Rev. Mod. Phys.* **56** (1984) 255.
8. R. A. Bernstein *et al.*, *Proc. SPIE-Int Soc. Opt. Eng.* **9145** (2014) 91451C.
9. P. Birch, *Nature* **298** (1982) 451.
10. A. H. Bridle and R. A. Perley, *Annu. Rev. Astron. Astrophys.* **22** (1984) 319.
11. M. L. Brown *et al.*, *Astrophys. J.* **705** (2009) 978.
12. S. M. Carroll, G. B. Field and R. Jackiw, *Phys. Rev. D* **41** (1990) 1231.
13. S. M. Carroll and G. B. Field, *Phys. Rev. Lett.* **79** (1997) 2394.
14. S. M. Carroll, *Phys. Rev. Lett.* **81** (1998) 3067.
15. S.-J. Chen, H.-H. Mei and W.-T. Ni, *Mod. Phys. Lett. A* **22** (2007) 2815.
16. A. Cimatti, S. di Serego Alighieri, R. A. E. Fosbury, M. Salvati and T. Duncan, *Mon. Not. Roy. Astron. Soc.* **264** (1993) 421.
17. A. Cimatti, S. di Serego Alighieri, G. B. Field and R. A. E. Fosbury, *Astrophys. J.* **422** (1994) 562.
18. J. N. Clarke, P. P. Kronberg and M. Simard-Normandin, *Mon. Not. Roy. Astron. Soc.* **190** (1980) 205.
19. E. Costa *et al.*, *Exp. Astron.* **28** (2010) 137.
20. S. Covino *et al.*, *Astron. Astrophys.* **348** (1999) L1.
21. T. de Zeeuw, R. Tamai and J. Liske, *The Messenger* **158** (2014) 3.
22. A. J. Dean *et al.*, *Science* **321** (2008) 1183.
23. S. di Serego Alighieri, A. Cimatti and R. A. E. Fosbury, *Astrophys. J.* **431** (1994) 123.
24. S. di Serego Alighieri, G. B. Field and A. Cimatti, *Astron. Soc. Pac. Conf. Ser.* **80** (1995) 276.
25. S. di Serego Alighieri, F. Finelli and M. Galaverni, *Astrophys. J.* **715** (2010) 33.
26. S. di Serego Alighieri, W.-T. Ni and W.-P. Pan, *Astrophys. J.* **792** (2014) 35.
27. D. J. Eisenstein and E. F. Bunn, *Phys. Rev. Lett.* **79** (1997) 1957.
28. M. Galaverni, G. Gubitosi, F. Paci and F. Finelli, *J. Cosmol. Astropart. Phys.* **8** (2015) 31.
29. M. Goldhaber and V. Trimble, *J. Astrophys. Astron.* **17** (1996) 17.
30. K. M. Gorski *et al.*, *Astrophys. J.* **622** (2005) 759.
31. D. Götz *et al.*, *Mon. Not. Roy. Astron. Soc.* **444** (2014) 2776.
32. G. Gubitosi *et al.*, *J. Cosmol. Astropart. Phys.* **0908** (2009) 021.
33. WMAP Collab. (G. Hinshaw *et al.*), *Astrophys. J. Suppl.* **208** (2013) 19.
34. D. Hutsemekers, R. Cabanac, H. Lamy and D. Sluse, *Astron. Astrophys.* **441** (2005) 915.
35. D. Hutsemekers, L. Braibant, V. Pelgrims and D. Sluse, *Astron. Astrophys.* **572** (2014) A18.
36. IAU Commission 40, *Trans. Int. Astron. Union* **XVB** (1974) 166.
37. S. A. Joshi, R. A. Battye, I. W. A. Browne, N. Jackson, T. W. B. Muxlow and P. N. Wilkinson, *Mon. Not. Roy. Astron. Soc.* **380** (2007) 162.
38. E. Kalemci *et al.*, *Astrophys. J. Suppl.* **169** (2007) 75.
39. M. Kamionkowski, *Phys. Rev. D* **82** (2010) 047302.
40. BICEP1 Collab. (J. P. Kaufman *et al.*), *Phys. Rev. D* **89** (2014) 062006.
41. J. P. Kaufman, B. G. Keating and B. R. Johnson, *Mon. Not. Roy. Astron. Soc.* **455** (2016) 1981.
42. B. G. Keating, M. Shimon and A. P. S. Yadav, *Astrophys. J. Lett.* **762** (2013) L23.
43. V. A. Kostelecký and M. Mewes, *Phys. Rev. Lett.* **87** (2001) 251304.
44. V. A. Kostelecký and M. Mewes, *Phys. Rev. D* **66** (2002) 056005.
45. V. A. Kostelecký and M. Mewes, *Phys. Rev. Lett.* **110** (2013) 201601.
46. J. M. Kovac, *et al.*, *Nature* **420** (2002) 722.

47. P. P. Kronberg, C. C. Dyer and H.-J. Röser, *Astrophys. J.* **472** (1996) 115.
48. J. P. Leahy, astro-ph/9704285.
49. N. F. Lepora, arXiv:gr-qc/9812077.
50. S. Liberati and L. Maccione, *Annu. Rev. Nucl. Part. Sci.* **59** (2009) 245.
51. T. J. Loredo, E. E. Flanagan and I. M. Wasserman, *Phys. Rev. D* **56** (1997) 7507.
52. L. Maccione, S. Liberati, A. Celotti, J. G. Kirk and P. Ubertini, *Phys. Rev. D* **78** (2008) 103003.
53. M. Massardi *et al.*, *Mon. Not. Roy. Astron. Soc.* **436** (2013) 2915.
54. H.-H. Mei, W.-T. Ni, W.-P. Pan, L. Xu and S. di Serego Alighieri, *Astrophys. J.* **805** (2015) 107.
55. R. C. Myers and M. Pospelov, *Phys. Rev. Lett.* **90** (2003) 211601.
56. S. Naess *et al.*, *J. Cosmol. Astropart. Phys.* **10** (2014) 007.
57. W.-T. Ni, A Nonmetric Theory of Gravity, Montana State University (1973), http://astrod.wikispaces.com/file/detail/A Non-metric Theory of Gravity.pdf.
58. W.-T. Ni, *Phys. Rev. Lett.* **38** (1977) 301.
59. W.-T. Ni, *Prog. Theor. Phys. Suppl.* **172** (2008) 49.
60. W.-T. Ni, *Rep. Prog. Phys.* **73** (2010) 056901.
61. W.-T. Ni, *Phys. Lett. A* **379** (2015) 1297.
62. W.-T. Ni, Equivalence Principles, Spacetime Structure and the Cosmic Connection, Chapter 5, this book; *Int. J. Mod. Phys. D* **25** (2016) 1630002.
63. B. Nodland and J. P. Ralston, *Phys. Rev. Lett.* **78** (1997) 3043.
64. L. Pagano *et al.*, *Phys. Rev. D* **80** (2009) 043522.
65. A. Payez, J. R. Cudell and D. Hutsemekers, *Phys. Rev. D* **84** (2011) 085029.
66. D. J. Saikia and C. J. Salter, *Annu. Rev. Astron. Astrophys.* **26** (1988) 93.
67. G. H. Sanders, *J. Astrophys. Astron.* **34** (2013) 81.
68. L. I. Schiff, *Am. J. Phys.* **28** (1960) 340.
69. L. Testi and J. Walsh, *The Messenger* **152** (2013) 2.
70. H. D. Tran, M. H. Cohen, P. M. Ogle, R. W. Goodrich and S. di Serego Alighieri, *Astrophys. J.* **500** (1998) 660.
71. J. Vernet, R. A. E. Fosbury, M. Villar-Martin, M. H. Cohen, A. Cimatti, S. di Serego Alighieri and R. W. Goodrich, *Astron. Astrophys.* **366** (2001) 7.
72. J. F. C. Wardle, R. A. Perley and M. H. Cohen, *Phys. Rev. Lett.* **79** (1997) 1801.
73. K. Wiersema *et al.*, *Nature* **509** (2014) 201.
74. J.-Q. Xia, H. Li, X. Wang and X. Zhang, *Astron. Astrophys.* **483** (2008) 715.
75. J.-Q. Xia, H. Li and X. Zhang, *Phys. Lett. B* **687** (2010) 129.
76. E. Zavattini *et al.*, *Phys. Rev. Lett.* **96** (2006) 110406.
77. E. Zavattini *et al.*, *Phys. Rev. D* **77** (2007) 032006.
78. N. Aghanim *et al.* (Planck Collaboration), submitted to *Astron. Astrophys.* (2016), arXiv:1605.08633.
79. W.-P. Pan, S. di Serego Alighieri, W.-T. Ni and L. Xu, to be published in *Proceedings of the Second LeCosPA Symposium "Everything about Gravity"*, December 14–18, 2015, Taipei (2016), arXiv:1603.08193.
80. P. de Bernardis and S. Masi, *Int. J. Mod. Phys. D* **25** (2016) 1640012.